\documentclass[aps,prl,twocolumn,showpacs,preprintnumbers,amsmath,amssymb]{revtex4}

\usepackage{graphicx}
\usepackage{dcolumn}
\usepackage{bm}

\begin{document}

\title{The complement: a solution to liquid drop finite size effects in phase transitions}

\author{L. G. Moretto$^1$, K. A. Bugaev$^1$, J. B. Elliott$^1$, R. Ghetti$^2$, J. Helgesson$^3$ and L. Phair$^1$}
\affiliation{$^1$Nuclear Science Division, Lawrence Berkeley National Laboratory,
Berkeley, CA  94720\\
$^2$Department of Physics, Lund University, Sweden\\
$^3$School of Technology and Society, Malm\"{o} University, Sweden}

\date{\today}
\begin{abstract}
The effects of the finite size of a liquid drop undergoing a phase transition are described in terms of the complement, the largest (but mesoscopic) drop representing the liquid in equilibrium with the vapor. Vapor cluster concentrations, pressure and density from fixed mean density lattice gas (Ising) calculations are explained in terms of the complement generalization of Fisher's model. Accounting for this finite size effect is important for extracting the infinite nuclear matter phase diagram from experimental data.
\end{abstract}

\preprint{LBNL-54448}
\maketitle

Finite size effects are essential in the study of nuclei and other mesoscopic systems for opposite, but complementary reasons. In modern cluster physics, the problem of finite size arises when attempts are made to relate known properties of the infinite system to cluster properties brought to light by experiment \cite{schmidt-01}. For nuclear physics, the problem is the opposite: finite size effects dominates the physics at all excitations and the challenge is generalize specific properties of a \emph{drop} (nucleus) to a description of uncharged, symmetric infinite nuclear matter. This goal has been achieved already for cold nuclei by the liquid drop model. Finite size effects are also encountered in nuclear physics in efforts to generate a liquid-vapor phase diagram from heat capacity measurements \cite{dagostino-00} and fragment distributions \cite{elliott-02}.

We present a general approach to deal with finite size effects in phase transitions and illustrate it for liquid-vapor phase coexistence. A dilute, nearly ideal vapor phase is in equilibrium with a denser liquid phase; finiteness is realized when liquid phase is a finite drop. A finite drop's vapor pressure is typically calculated by including the surface energy in the molar vaporization enthalpy \cite{rayleigh-17,moretto-02}.  We introduce here the concept of the complement to extend and quantify finite size effects down to drops as small as atomic nuclei. We generalize Fisher's model \cite{fisher-69}, deriving an expression for cluster concentrations of a vapor in equilibrium with a finite drop and recover from it the Gibbs-Thomson formulae \cite{krishnamachari-96}.  We demonstrate our approach with the lattice gas (Ising) model.

The complement method consists of evaluating the free energy change occurring when a cluster moves from one phase to another. For a finite liquid drop in equilibrium with its vapor, this is done by virtually transfering a cluster from the liquid drop to the vapor and evaluating the energy and entropy changes associated with \emph{both} the vapor cluster \emph{and} the residual liquid drop (complement). This method can be generalized to incorporate energy terms common in the nuclear case: symmetry, Coulomb (with caution \cite{moretto-03}) and angular momentum energies.

In the framework of physical cluster theories of non-ideal vapors (which assume the monomer-monomer interaction is exhausted by the formation of clusters), clusters behave ideally and are independent of each other. The complement method is based upon this independence. Physical cluster theories state that the concentrations of vapor clusters of $A$ constituents $n_A(T)$ depend on the free energy of cluster formation $\Delta G_A(T) = \Delta E_A(T) - T \Delta S_A(T)$. The epigon of physical cluster theories is Fisher's model \cite{fisher-69} which writes, at coexistence, $\Delta E = c_0 A^{\sigma}$ and $\Delta S_A(T) = \frac{c_0}{T_c}A^{\sigma} - \tau \ln A $. Thus
\begin{eqnarray}
	n_A(T)  =  \exp \left[-\frac{\Delta G_A(T)}{T} \right] = q_0A^{-\tau} \exp \left( - \frac{c_0 A^{\sigma} \varepsilon }{T} \right)
\label{eq:fisher-1}
\end{eqnarray} 
where $q_0$ is a normalization, $\tau$ is Fisher's topological exponent, $c_0$ is the surface energy coefficient, $\sigma$ is the surface to volume exponent, and $\varepsilon=(T_c-T)/T_c$. The leading term in $\Delta S_A(T)$, proportional to $A^{\sigma}$, permits the vanishing of the cluster free energy at a $T=T_c$ independent of size. Equation~(\ref{eq:fisher-1}) (and the extension below) is valid only at phase coexistence for $T \le T_c$. The direct physical interpretation of the parameters in $\Delta G_A(T)$ and their application to the nuclear case is the reason for this choice here, despite its limitations \cite{mader-03}.

We generalize eq.~(\ref{eq:fisher-1}), valid for infinite liquid-vapor equilibrium, to the case of a vapor in equilibrium with a finite liquid drop. For each vapor cluster we can perform the gedanken experiment of extracting it from the liquid, determining entropy and energy changes of the drop and cluster, and then putting it back into the liquid (the equilibrium condition), as if, according to physical cluster theories, no other clusters existed. Fisher's expressions for $\Delta E_A(T)$ and $\Delta S_A(T)$ can be written for a drop of size $A_{\rm d}$ in equilibrium with its vapor as $\Delta E_A(T) = c_0 \left[ A^{\sigma} + (A_{\rm d}-A)^{\sigma}-A_{\rm d}^{\sigma} \right]$ and $\Delta S_A(T) =  \frac{c_0}{T_c} \left[ A^{\sigma} + (A_{\rm d}-A)^{\sigma}-A_{\rm d}^{\sigma} \right]- \tau \ln \left[ {A(A_{\rm d}-A)}/{A_{\rm d}} \right]$ giving
\begin{eqnarray}
	n_A(T) & = & q_0\left[ \frac{A \left(A_{\rm d}-A\right)}{A_{\rm d}}\right]^{-\tau} \nonumber \\
	 & & \exp \left\{- \frac{c_0 \varepsilon }{T}\left[ A^{\sigma} + (A_{\rm d}-A)^{\sigma}-A_{\rm d}^{\sigma} \right] 
	 \right\}.
\label{eq:complement-2}
\end{eqnarray}
The free energy cost of complement ($A_{\rm d}-A$) formation is determined just as the free energy cost of cluster ($A$) formation is determined. The resulting expression Eq.~(\ref{eq:complement-2}) reduces to Eq.~(\ref{eq:fisher-1}) \emph{when $A_{\rm d}$ tends to infinity} and contains exactly the same parameters. We can rewrite Eq.~(\ref{eq:complement-2}) as
	\begin{equation}
	n_A(T) = n_{A}^{\infty}(T) \exp \left( \frac{A\Delta \mu_{\text{fs}}}{T} \right) 
	\label{fisher2}
	\end{equation} 
with $n_{A}^{\infty}(T)$ given by Eq.~(\ref{eq:fisher-1}). The finite size of the drop acts as an {\it effective} chemical potential, $\Delta \mu_{\rm fs} = - \left\{  c_0 \varepsilon \left[ \left(A_{\rm d}-A \right)^{\sigma}-A_{\rm d}^{\sigma} \right] - T \tau \ln\left[ \left( A_{\rm d}-A\right) / A_{\rm d} \right] \right\} / A$, increasing the vapor pressure \cite{krishnamachari-96}.

\begin{figure}
\includegraphics[width=8.7cm]{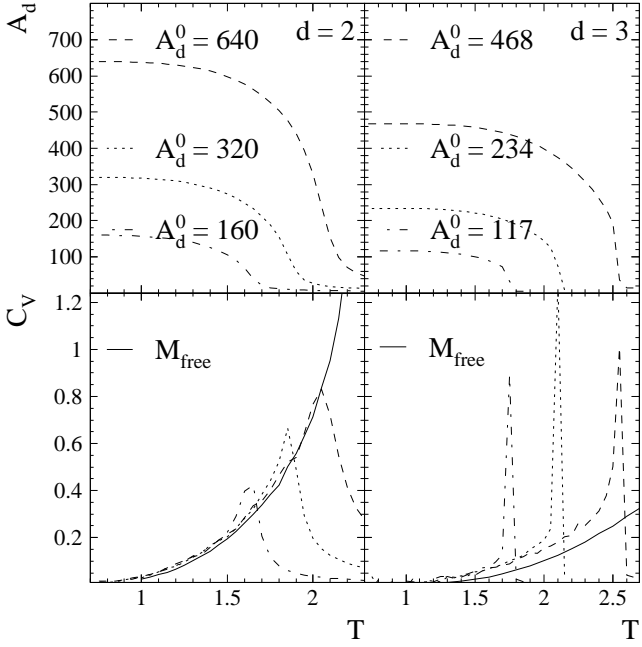}
\caption{On the left (right), top to bottom: the $d=2$ ($3$) liquid drop size $A_{\rm d}$, specific heat $C_V$. See text for details.}
\label{systematics}
\end{figure}

In order to quantitatively demonstrate this method, we apply it to the canonical lattice gas (Ising) model \cite{lee-52.1,lee-52.2} with a fixed number of up spins, i.e. a fixed mean occupation density $\rho_{\rm fixed}$ lattice gas (equivalently, a fixed magnetization $M_{\rm fixed}$ Ising model) \cite{elliott-04}. Up spins represent particles of the fluid forming monomers, dimers, drops etc. Down spins are empty space; the lattice is the container enclosing the fluid. We chose large lattices with periodic boundary conditions to minimize finite lattice effects, irrelevant to our study.  For $d=2$ ($3$) we used a square (simple cubic) lattice of side $L=80$ ($25$) which leads to a shift in $T_c$ of $\lesssim 0.5$\% \cite{ferdinand-69,landau-76.1} ($\lesssim 0.5$\% \cite{landau-76.2,ferrenberg-91}). The $M_{\rm fixed}$ calculations were performed according to ref. \cite{heringa-98}. For every $(T,{\rho}_{\rm fixed})$ over $10^5$ thermalized realizations were generated to produce the cluster concentrations. We performed $M_{\rm free}$ calculations for the same lattices as a benchmark in order to differentiate effects of a finite lattice from those of a finite drop.

For the $M_{\rm fixed}$ calculations at $T=0$, the up spins aggregate into a single liquid drop in a vacuum: the ground state. At higher temperatures, the vacuum is filled with a vapor made of up spin clusters. Clusters in the vapor were identified via the Coniglio-Klein algorithm \cite{coniglio-80} to insure that their behavior is physical (i.e. cluster concentrations return Ising critical exponents and not percolation exponents). The largest cluster represents the liquid drop and is identified geometrically (all like spin nearest neighbors bonded) in order to capture the skin thickness associated with liquid drops \cite{elliott-04}. Our choices of ground state liquid drops $A_{\rm d}^{0}$ (shown in Fig.\ref{systematics}) insure that the ground state is approximately square (cubical) for $d=2$ ($3$). Due to periodic boundary conditions the ground state shape changes with $M_{\rm fixed}$ \cite{binder-81}.

\begin{figure}
\includegraphics[width=8.7cm]{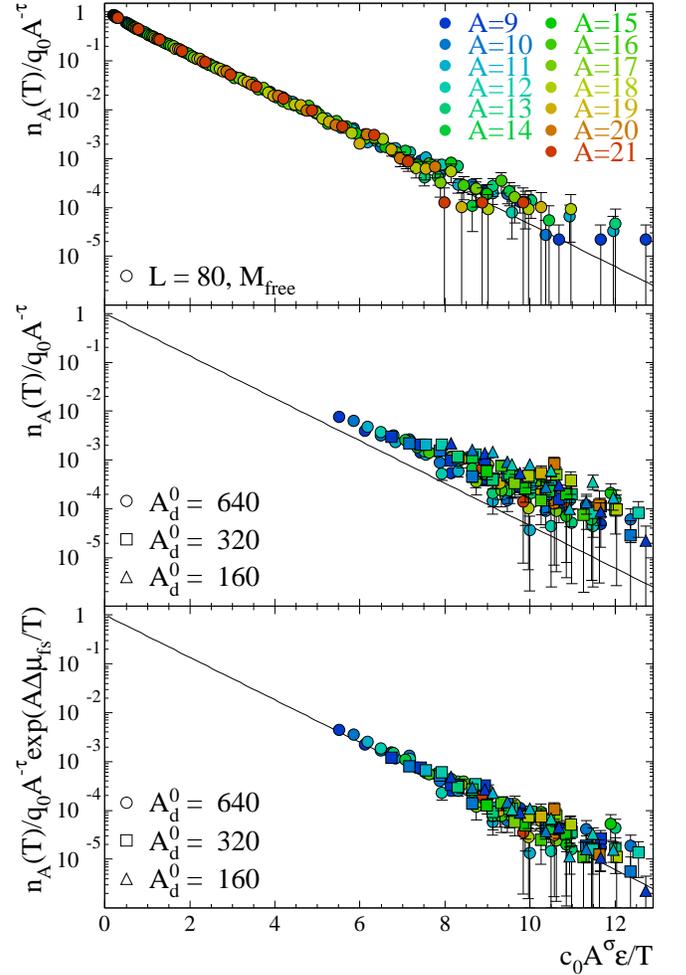}
\caption{[Color online] The cluster concentrations of the $d=2$, $L=80$ periodic boundary condition square lattice for: $M_{\rm free}$ calculations (top); $M_{\rm fixed}$ calculations with no complement (middle); $M_{\rm fixed}$ calculations with the complement (bottom).}
\label{cluster-yields-figure}
\end{figure}

\begin{figure}
\begin{center}
\includegraphics[width=8.7cm]{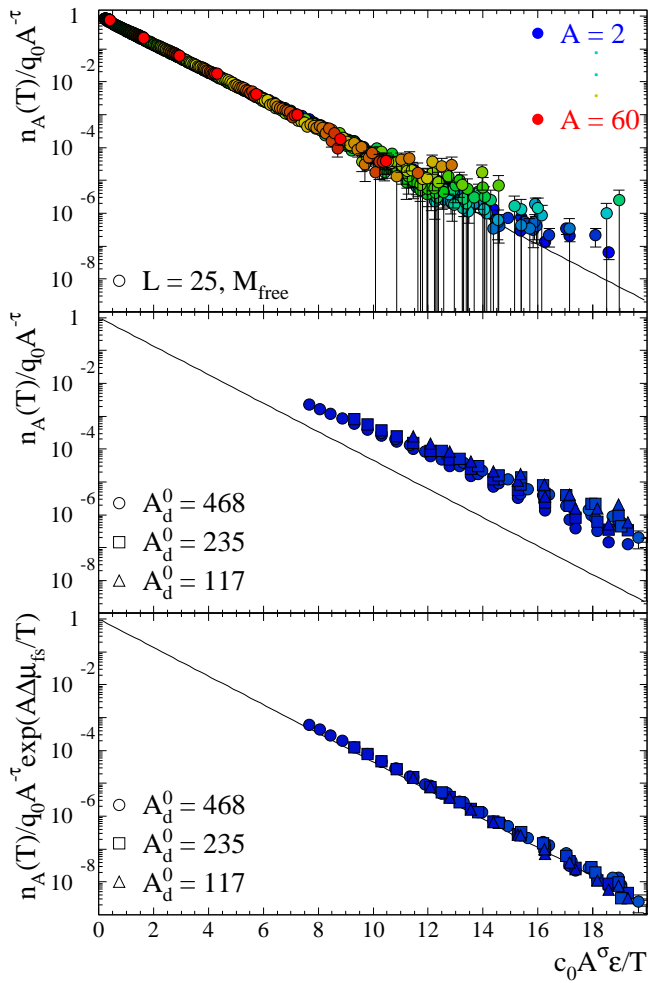}
\caption{[Color online] Same as Fig.~\ref{cluster-yields-figure} but for the $d=3$, $L=25$  periodic boundary condition simple cubic lattice.}
\label{fig:$d=3$}
\end{center}
\end{figure}

Figure~\ref{systematics} shows that as $T$ increases, the drop's size $A_{\rm d}$ decreases from $A_{\rm d}^{0}$; the evaporating drop fills the container with vapor. At a temperature $T_X$, corresponding to the end of two phase coexistence, $A_{\rm d}$ falls quickly. The value of $T_X$ varies with $A_{\rm d}^{0}$. For $T \lesssim T_X$ the specific heat $C_V$ (measuring spin-spin interaction energy fluctuations only) agrees approximately with the $M_{\rm free}$ results (solid lines in Fig.~\ref{systematics}) for all $A_{\rm d}^{0}$ until fluctuations in $A_{\rm d}$ at $\sim T_X$ produce a  $C_V$ greater than that of the $M_{\rm free}$ results. As $T$ increases further, $C_V$ decreases.

To evaluate the efficacy of the complement, we examine the scaled cluster concentrations for our calculations:  $n_A(T)/q_0 A^{-\tau}$ vs. $c_0 A^{\sigma} \varepsilon / T$. For $M_{\rm free}$ calculations it has been shown that this scaling collapses concentrations of clusters over a wide range in $A$ and $T$ \cite{mader-03}. Finite size liquid drop effects will be manifested in the cluster concentrations of the $M_{\rm fixed}$ calculations which should scale better according to Eq.~(\ref{eq:complement-2}) than to Eq.~(\ref{eq:fisher-1}).

\begin{table}[htdp]
\caption{Fit results for $M_{\rm free}$ calculations}
\begin{center}
\begin{tabular}{ccccc}
\hline
          -                             &     Onsager & this work       & theoretical values                      & this work         \\
          -                             &    $d=2$ & $d=2$        & $d=3$                      & $d=3$ \\
          -                             &     $L \rightarrow \infty$ & $L=80$        & $L \rightarrow \infty$                      & $L=25$         \\
        ${\chi}^{2}_{\nu}$ &          -               & $4.7$                         & -                                          & $1972.2$                       \\
        $T_c$                     & $2.26915$      & $2.283 \pm 0.004$ & $4.51152\pm0.00004$  & $4.533 \pm 0.002$  \\
        $c_0$                     &        $\ge8$         & $8.6 \pm 0.2$           & $\ge12$                               & $12.63 \pm 0.04$      \\
        $\sigma$                &    $8/15$         & $0.56\pm0.01$    & $0.63946\pm0.0008$    & $0.725 \pm 0.003$ \\
        $\tau$                     &    $31/15$       & $2.071\pm0.002$    & $2.209\pm0.006$           & $2.255 \pm 0.001$ \\
\hline
\end{tabular}
\end{center}
\label{free}
\end{table}

\begin{table}[htdp]
\caption{${\chi}^{2}_{\nu}$ results for $M_{\rm fixed}$ calculations}
\begin{center}
\begin{tabular}{cccc}
\hline
        $A_{\rm d}^{0}$ ($d=2$)                                    & $640$   &$320$   & $160$  \\
        ${\chi}^{2}_{\nu}$ w/o complement & $10.3$ & $10.6$ & $18.2$  \\
        ${\chi}^{2}_{\nu}$ w complement   & $1.7$    & $1.9$   & $4.8$\\
\hline
         $A_{\rm d}^{0}$ ($d=3$) &$468$    & $234$       &$117$ \\
        ${\chi}^{2}_{\nu}$ w/o complement &$14~825.9$ & $7~938.6$ & $3~516.0$\\
         ${\chi}^{2}_{\nu}$ w complement &$1~553.4$     & $838.6$    & $258.1$ \\
\hline
\end{tabular}
\end{center}
\label{fixed}
\end{table}

Only clusters of $A\ge9$ are included in the $M_{\rm free}$ fits for the $d=2$ calculations. This is because only large clusters obey Fisher's ansatz for the cluster surface energy: $E_A = c_0 A^{\sigma}$. Small clusters are dominated by geometrical shell effects \cite{elliott-03}. For the $d=3$ $M_{\rm fixed}$ calculations, large clusters are very rare, so clusters of $A\ge2$ are included in our analysis. Thus, the magnitudes of the $\chi^{2}_{\nu}$ values from the $d=3$ calculations are due to the clusters analyzed not following closely Fisher's ansatz.

In the thermodynamic limit, the highest temperature admitted by Eq.~(\ref{eq:fisher-1}) or (\ref{eq:complement-2}) is the temperature at which the system leaves coexistence.  For the $M_{\rm free}$ calculations this occurs at $T=T_c$, while for the $M_{\rm fixed}$ calculations this occurs at $T \approx T_X$. However, due to the small size of our drops, fluctuations grow large before $T_X$ and we consider only $T \lesssim 0.85T_X$ ($0.75T_X$) for $d=2$ ($3$).

To make a comparison between the scaling achieved with the $M_{\rm free}$ clusters and with the $M_{\rm fixed}$ clusters, we fit the $M_{\rm free}$ clusters with Eq.~(\ref{eq:fisher-1}) with the free parameters $T_c$, $c_0$, $\sigma$ and $\tau$; $q_0=\zeta(\tau-1)/2$. Results are given in Table~\ref{free} and shown in top panels of figures~\ref{cluster-yields-figure} and \ref{fig:$d=3$}. The values of the $T_c$ and $\tau$ returned by this procedure are within $1$\% of their established values. The value of $c_0$ shows that all clusters are not perfect squares (cubes) in $d=2$ ($3$) for which $c_0=8$ ($12$) \cite{mader-03,stauffer-99}.  The values of $\sigma$ are within $5\%$ ($15\%$) of their established values for $d=2$ ($3$). This level of inaccuracy arises from shell effects \cite{elliott-03}.

Next we calculate $\chi^{2}_{\nu}$ for the $M_{\rm fixed}$ clusters using Eq.~(\ref{eq:fisher-1}) and Eq.~(\ref{eq:complement-2}) with parameters fixed to the Table~\ref{free} values for the infinite system. This procedure frees us as much as possible from the drawbacks of Fisher's model so we can concentrate on the effect of the complement. Results are given in Table~\ref{fixed} and shown in the middle panels (without complement: concentrations scaled as $n_A(T)/q_0 A^{-\tau}$) and the bottom panels (with complement: concentrations scaled as $n_A(T)/q_0 A^{-\tau}\exp\left[A\Delta \mu_{\rm fs}/T\right]$) of figures~\ref{cluster-yields-figure} and \ref{fig:$d=3$}. The $M_{\rm fixed}$ $\chi^{2}_{\nu}$ values for the calculation with the complement are an order of magnitude smaller than the results without the complement and the data collapse is better.

\begin{figure}
\includegraphics[width=8.7cm]{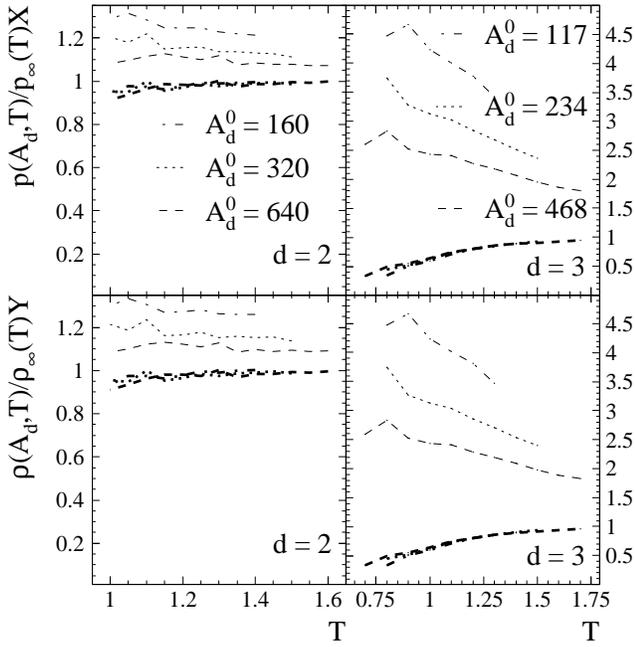}
\caption{ Left (right) the normalized pressure and density of a vapor in coexistence with a drop $A_{\rm d}$ for $d=2$ ($3$). Thin [thick] lines show no complement results, $X=Y=1$ [complement results,  $X$ and $Y$  from Eqs.~(\ref{pressure}) and (\ref{density})].}
\label{fig:pressure}
\end{figure}

The middle and bottom panels of figures~\ref{cluster-yields-figure} and \ref{fig:$d=3$} show the major point of this paper: generalizing Fisher's model with the complement accounts for the finiteness of the liquid. The middle panels of figures~\ref{cluster-yields-figure} and \ref{fig:$d=3$} show that the $M_{\rm fixed}$ calculations do not scale as their $M_{\rm free}$ counterparts. The bottom panels of figures~\ref{cluster-yields-figure} and \ref{fig:$d=3$} show that when the complement effect is taken into account, the $M_{\rm fixed}$ calculations scale as their $M_{\rm free}$ counterparts.

We now confront the integrated quantities, pressure $p(A_{\rm d},T) = T \sum_A n_A(T)$ and density ${\rho}(A_{\rm d},T) = \sum_A n_A(T)A$ . For $A_{\rm d} \gg A$, expanding $\Delta \mu_{\rm fs}$ gives
	\begin{equation}
	\Delta \mu_{\rm fs} = 1 + A\left( \frac{\tau}{A_{\rm d}} + \frac{\sigma c_0 \varepsilon}{TA_{\rm d}^{1-\sigma}} \right) + \cdots
	\label{expansion}
	\end{equation}
This leads to
	\begin{eqnarray}
	p(A_{\rm d},T) & \approx&  p_{\infty}(T) \exp \left[ \left( \frac{\tau}{A_{\rm d}} + \frac{\sigma c_0 \varepsilon}{TA_{\rm d}^{1-\sigma}} \right) \frac{\sum_{A} n_{A}^{\infty}(T) A}{\sum_{A} n_{A}^{\infty}(T)} \right] \nonumber \\
	& = & p_{\infty}(T) X
	\label{pressure}
	\end{eqnarray}
and
	\begin{eqnarray}
	{\rho}(A_{\rm d},T) & \approx & {\rho}_{\infty}(T) \exp \left[ \left( \frac{\tau}{A_{\rm d}} + \frac{\sigma c_0 \varepsilon}{TA_{\rm d}^{1-\sigma}} \right) \frac{\sum_A n_{A}^{\infty}(T) A^2}{\sum_A n_{A}^{\infty}(T)A} \right] \nonumber \\
	& = & {\rho}_{\infty}(T) Y.
	\label{density}
	\end{eqnarray}
For a vapor of monomers as $A_{\rm d} \gg \tau$ equations~(\ref{pressure}) and (\ref{density}) yield the Gibbs-Thomson formulae \cite{krishnamachari-96}.

Figure~\ref{fig:pressure} shows the behavior of $p(A_{\rm d},T)$ and ${\rho}(A_{\rm d},T)$ for the $M_{\rm fixed}$ calculations compared to the bulk results. To free ourselves from finite lattice size effects $p_{\infty}(T)$, ${\rho}_{\infty}(T)$ and $n_{A}^{\infty}(T)$ are determined from the $M_{\rm free}$ calculations. As expected $p(A_{\rm d},T)>p_{\infty}(T)$ and ${\rho}(A_{\rm d},T)>{\rho}_{\infty}(T)$ (thin lines in Fig.~\ref{fig:pressure}); i.e. the ratio in question is $>1$. Accounting for the complement via equations~(\ref{pressure}) and (\ref{density}) (using values in Table~\ref{free}) collapses results from all the calculations to a single line recovering the bulk behavior (thick lines in Fig.~\ref{fig:pressure}); i.e. the ratio in question is $\sim 1$. Deviations at low $T$ are due to the increasing effects of monomers which have $c_0=8$ ($12$) in $d=2$ ($3$).

In conclusion, a general approach in terms of the complement has been developed which allows one to account for finite liquid drop size effects and to extrapolate from finite to infinite systems.  We have demonstrated the applicability of our method using lattice gas model (Ising) calculations. This method can be generalized to include other energy factors present in the nuclear case (e.g. symmetry, Coulomb and angular momentum energies) which is another important step towards determining the liquid-vapor phase boundary of infinite, symmetric nuclear matter from experimental nuclear data.

\end{document}